\documentclass[]{aastex631}

\shorttitle{Solar calibration of $\alpha_{\rm MLT}$ using {\AE}SOPUS opacities in MESA}
\shortauthors{Cinquegrana and Joyce}

\graphicspath{{./}{figures/}}

\usepackage{listings}

\begin{document}

\title{Solar calibration of the convective mixing length for use with the {\AE}SOPUS opacities in MESA}

\author[0000-0001-7902-8134]{Giulia C. Cinquegrana}
\affiliation{School of Physics \& Astronomy, Monash University, Clayton VIC 3800, Australia\\
}
\affiliation{ARC Centre of Excellence for All Sky Astrophysics in 3 Dimensions (ASTRO 3D) \\
}

\author[0000-0002-8717-127X]{Meridith Joyce}
\affiliation{Space Telescope Science Institute, 3700 San Martin Drive, Baltimore, MD 21218, USA\\
}
\affiliation{ARC Centre of Excellence for All Sky Astrophysics in 3 Dimensions (ASTRO 3D) \\
}

\begin{abstract}
The simplistic but ubiquitous Mixing Length Theory (MLT) formalism is used to model convective energy transport within 1D stellar evolution calculations. The formalism relies on the free parameter $\alpha_{\rm MLT}$, which must be independently calibrated within each stellar evolution program and for any given set of physical assumptions. We present a solar calibration of $\alpha_{\text{MLT}}$ appropriate for use with the {\AE}SOPUS opacities, which have recently been made available for use with the MESA stellar evolution software. We report a calibrated value of $\alpha_{\rm MLT}=1.931$ and demonstrate the impact of using an appropriately calibrated value in simulations of a $3 M_{\odot}$ asymptotic giant branch star. 
\end{abstract}

\section{Introduction} 
\label{introduction}

The Mixing Length Theory of convection (MLT) was first applied to stellar interiors by Erika Boehm-Vitense \citep{Vitense53} and remains one of few we have for approximating convection in 1D stellar models on evolutionary timescales (along with, e.g., the Full Spectrum of Turbulence model of \citealt{Canuto91, Canuto96}). The MLT formalism contains a free parameter, $\alpha_{\text{MLT}}$, that can be thought of as characterizing the ``efficiency of convection,'' or the ``mean-free path'' of a fluid parcel by analogy with molecular heat transfer \citep{Prandtl25}. As $\alpha_{\text{MLT}}$ does not have a physical meaning in 3D fluid dynamics, it must be calibrated empirically. Most often, and for obvious reasons, this calibration is performed to the Sun \citep[though exceptions include][]{Joyce18not, Joyce18class}. However, as $\alpha_{\text{MLT}}$ may absorb modeling inconsistencies or inherit other artifacts from the environment in which it is used, $\alpha_{\rm MLT}$ must be calibrated anew for each stellar evolution code and for each choice of input physics--especially when the physical assumptions heavily impact the outer layers of stars: regions where the temperature gradient becomes superadiabatic (see, e.g., \citealt{Jermyn2022} for detailed discussion). Different choices of opacities, for example, will therefore require different solar mixing length calibrations.

\section{Methods}
\label{}
This note is a tangential result from the work of Cinquegrana, Joyce and Karakas (2022, \textit{\textit{in prep}}), wherein MESA is used to model high- (i.e. super-solar) metallicity intermediate mass and massive stars. We discuss here the relevant physics for this science case as well as the variations applied for the current demonstration.

\subsection{Software tools}
The {\AE}SOPUS: Accurate Equation of State and OPacity Utility Software \citep{Marigo09} tool allows users to create low-temperature Rosseland mean opacity tables customized to the situation they are modelling by, e.g., choosing different CNO abundances or alpha-element enhancement levels. The necessity of using low-temperature opacity tables that can follow composition changes along the evolution of a star, rather than relying on the star's initial metal content, has been reiterated by numerous groups over the past two decades \citep[e.g.,][]{Marigo02opacities, Cristallo07molec, Ventura09evo, Constantino14nec, Fishlock14opacities}. 
The importance of this feature is especially apparent when modeling thermally pulsing asymptotic giant branch (AGB) stars, which undergo various processes such as third dredge up and hot bottom burning that significantly change the chemical composition of the envelope from its initial chemical makeup \citep[for further details on AGB stars, see:][]{Busso99, Herwig05,  Nomoto13, Karakas14, Karakas22paper1, Cinquegrana22paper2, Ventura22nuc}.
Such opacity tables have also proved useful in modelling R Coronae Borealis stars, which have envelopes deficient of hydrogen but enhanced in CNO elements \citep{Schwab19}. The impacts of using appropriate opacities in these situations further extend to stellar yield calculations as well as predictions for envelope expansion, mass loss rates and stellar lifetimes. 

The {\AE}SOPUS tool has been available since 2009, however the integration of {\AE}SOPUS with the Modules for Experiments in Stellar Astrophysics (MESA; \citealt{Paxton10instrument1, Paxton13instrument2, Paxton15instrument3, Paxton18instrument4, Paxton19instrument5})
software is recent (MESA Instrument Paper VI, \textit{in prep}). \citet{Schwab19} was the first to publish science utilising {\AE}SOPUS within MESA, however, there is currently no commentary on the appropriate mixing length to use with these opacities published in the literature.

\subsection{Physical configuration} 

We perform a solar calibration of $\alpha_{\rm MLT}$ using MESA version 15140 \citep{Paxton19instrument5}. Opacities are scaled according to the abundances of \citet{Lodders03}. We determine the best-fitting mixing length by comparing a number of solar models, each adopting a different mixing length and initial helium abundance, $Y_{\rm initial}$. All models are assigned $Z_{\rm initial} = 0.0133$ per the solar metallicity determination of \citet{Lodders03} for consistency with the {\AE}SOPUS opacity assumptions.

Initial helium abundances from 0.2377 to 0.28 are considered (for discussion of varying $Y_{\rm initial}$ as part of mixing length calibrations, see \citet{Joyce18not}, \citet{Trampedach2014}, and references therein). 
The base metal content for the opacity tables, \verb|Zbase|, is set equal to $Z_{\rm initial}$. OPAL data are used for high temperature opacities \citep{Iglesias96}; low temperature opacities are built using the {\AE}SOPUS tool \citep{Marigo09}.
\footnote{The initial conditions used to create our tables can be found in the appendix \S~\ref{aesopusinstructions}.
%
Instructions on how to prepare the tables for use within MESA can be found within the directory: mesa-r15140/kap/preprocessor/AESOPUS/README} 

All models used for the solar calibration have $1 M_{\odot}$ and terminate at the solar age, but are otherwise identical to the physical prescription described above and adopted in Cinquegrana, Joyce \& Karakas (2022, \textit{in prep}). Using the MLT prescription of \citet{Henyey65mlt}, values are iterated until some combination of $Y_{\rm initial}$ and $\alpha_{\rm MLT}$ reproduces the solar radius and solar luminosity at the solar age to approximately 1 part in $10^4$.

Reference values for solar fundamental parameters used in this work are as follows:
\begin{itemize}
\item $t_{\odot} = 4.57 \times 10^9$ years \citep{Bahcall95solar};
\item $L_{\odot} = 3.846 \times 10^{33}$ erg s$^{-1}$ \citep{Willson97total}; and
\item $R_{\odot} =  6.95508 \times 10^{10}$ cm \citep{Brown98accurate}. 
\end{itemize}
These are collated in \citet{Christensen-Dalsgaard21solar}.

\section{Results} 
\begin{figure*} 
	\includegraphics[width=18cm]{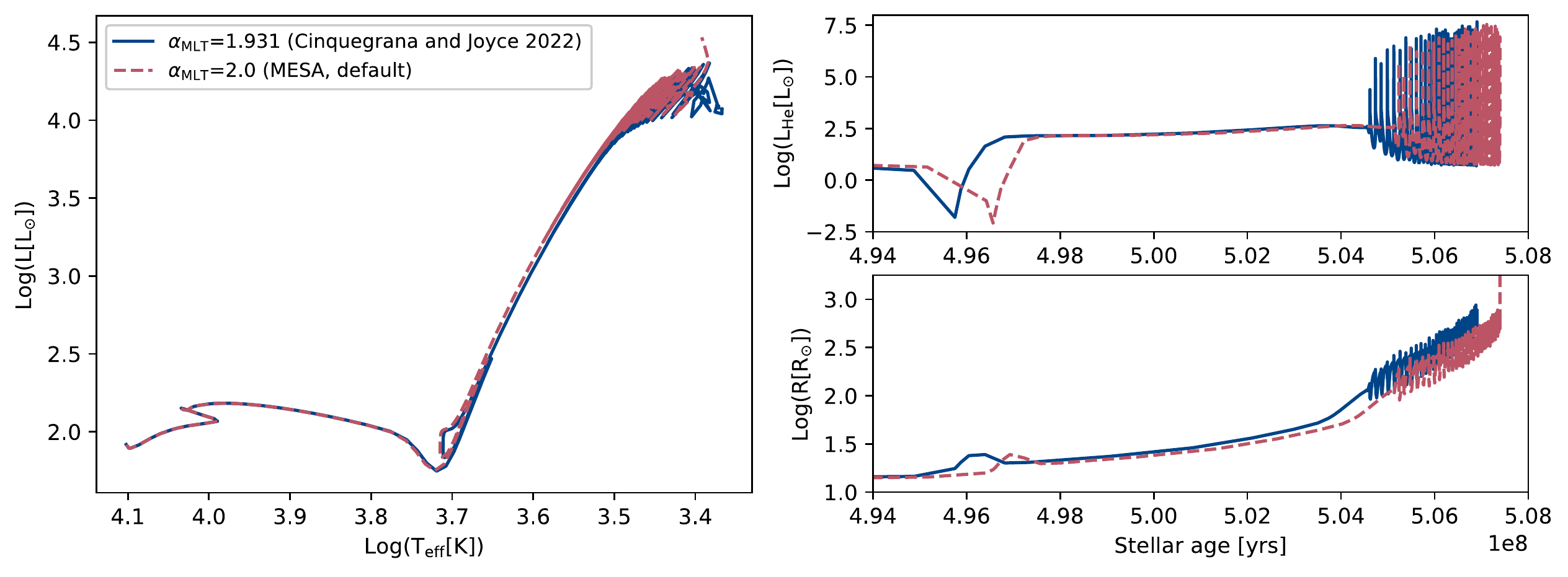}
    \caption{Two 3\(M_\odot\), solar metallicity models are compared: one with $\alpha_{\rm MLT}$=1.931 (continuous line) and one with  $\alpha_{\text{MLT}} = 2.0$ (dashed line). The left panel shows the stellar tracks of the two models.
    The right panels shows the variance of the helium burning luminosity and radius as the models approach the thermally pulsing AGB.}
    \label{fig:3Mcomp}
\end{figure*}

For each composition assumption, we run a suite of models adopting $\alpha_{\rm MLT}$ values between 1.3 and 2.2. We extract the luminosity and radius at the solar age, $t_{\odot}$. We then determine which choice of $\alpha_{\text{MLT}}$ produces the best reproduction of solar parameters according to a penalty statistic that prioritizes agreement with the solar radius and solar luminosity equally. The precision on $\alpha_{\rm MLT}$ is increased until further refinement no longer reduces the residuals in luminosity and radius. We find that a combination of $Y_{\rm initial}=0.241$ and $\alpha_{\rm MLT} = 1.931$ performs best. 

\section{Implications}

Composition-dependent opacity tables, such as those provided by {\AE}SOPUS, are crucial when modelling AGB stars. The reliability of AGB stellar models therefore depends on using a value of $\alpha_{\rm MLT}$ that is appropriately calibrated for use with those opacities.
In particular, the mixing length is known to have a large impact on derived quantities such as the effective temperature and radius \citep{Karakas14, Joyce20standing}, the efficiency of the third dredge up \citep{Boothroyd88}, and the strength of hot bottom burning \citep{Ventura05}.
Using an inappropriate $\alpha_{\rm MLT}$ can therefore lead to inaccurate predictions for fundamental stellar parameters.

This is demonstrated in Fig.\ \ref{fig:3Mcomp}, which compares the evolutionary properties of two 3\(M_\odot\) ($Z=$\(Z_\odot\)) models: one with $\alpha_{\rm MLT}=1.931$ and one with MESA's default value of  $\alpha_{\text{MLT}} = 2.0$. In the left panel, we note little difference in their tracks along the main sequence, though  variance in luminosity and temperature is apparent once the models ascend their giant branches. However, when evolved until the hydrogen envelope is less than 20\% of the total stellar mass, (e.g., the post-AGB EEP as defined in \citet{Dotter16eeps}),
we note that the $\alpha_{\text{MLT}} = 2.0$ model has experienced 35 thermal pulses by this point, compared to the  36 thermal pulses of the model with $\alpha_{\text{MLT}} = 1.931$. As the models evolve along the thermally pulsating AGB, we also see modest variation in radius, effective temperature and the magnitude of the helium burning luminosity. 

More striking, however, is the impact of decreasing $\alpha_{\rm MLT}$ on the evolutionary timescales. The model that assumes a calibrated $\alpha_{\rm MLT}$ value produces an earlier second dredge up event and shows an earlier onset of thermal pulses. Changes in timescale carry implications for stellar lifetimes and thus for nucleosynthesis and chemical evolution timescales.

In light of the clear physical changes imparted to the models by a seemingly modest absolute change in $\alpha_{\rm MLT}$, we provide a working ``best practices'' value of  $\alpha_{\text{MLT}} = 1.931$ for use with {\AE}SOPUS low-temperature opacities.

\section{Acknowledgements}
G.C.\ thanks A. Karakas for fruitful discussion of the literature and acknowledges the Australian Research Council Centre of Excellence for All Sky Astrophysics in 3 Dimensions (ASTRO 3D), project number CE170100013.
M.J.\ acknowledges the Lasker Data Science Fellowship, awarded by the Space Telescope Science Institute.

\appendix
\section{Initial conditions for {\AE}SOPUS tables}
\label{aesopusinstructions}

The {\AE}SOPUS software is available online at \url{http://stev.oapd.inaf.it/cgi-bin/aesopus}. The tables used in this work were computed using the following options.

\subsection{Temperature and density grid} 
We set logT from 3.2 to 4.5. Increments of 0.01 dex are used from 3.2 to 3.7; 0.05 dex is used above this. Similarly, logR extends from $-7$ to 1 in increments of 0.5 dex. 

\subsection{Chemical composition}
We use the reference solar composition of \citet{Lodders03}. We solve for reference metallicities of 0.014 to 0.10, given the high metallicities required for our science case. We set hydrogen abundance, X, from 0.4 to 0.8 in steps of 0.1. We do not change the abundance normalization and primordial mixture options. For the reference mixture, we choose \verb|scaled-solar|, but leave the associated table uncompleted. We do not define any additional enhancement or depletion factors in the \verb|superimposed chemical pattern| table.

The last option is for CNO abundance variation factors, where \verb|fc| and \verb|fn|
are scaling factors applied to the reference abundance (see \citet{Marigo09, Fishlock14opacities} for more detailed discussion). We note that the following values depend heavily on the mass, metallicity and evolutionary stage under consideration. In our case, none of our AGB models with 0.04 $\geq$ \verb|Zref| $\geq$ 0.10 become carbon rich (C/O $>$ 1), so we do not account for \verb|fc| values greater than 0.6. Solar metallicity AGB models, on the other hand, are very capable of exceeding this threshold, so in that case we extend \verb|fc| $>1$. For 0.04 $\geq$ \verb|Zref| $\geq$ 0.10, we set \verb|fc| = $-0.2$, 0, 0.2, 0.4, 0.6 and \verb|fn| = 0, 0.4, 0.7. For \verb|Zref|~= 0.014, we set \verb|fc| = 0, 0.2, 0.4, 0.6, 1.0, 1.5 and \verb|fn| = 0, 0.4, 0.7.

\bibliography{Giulias_bib}{}
\bibliographystyle{aasjournal}

\end{document}